# Using an instrumented hammer during Summers osteotomy: an animal model

Yasuhiro Homma[1,2,3], Manon Bas Dit Nugues[1], Arnaud Dubory [4,5], Charles-Henri Flouzat-Lachaniette[4,5], Jean-Paul Meningaud [4], Barbara Hersant [4], Emmanuel Gouet [6], Guillaume Haïat[1]*.

1. CNRS, Laboratoire Modelisation et Simulation Multi Echelle, MSME UMR 8208 CNRS, 61 Avenue du General de Gaulle, 94010 Creteil, France.
2. Department of Medicine for Orthopaedics and Motor Organ, Juntendo University Graduate School of Medicine, 2-1-1, Hongo, Bunkyo-ku, Tokyo 113-0033, Japan
3. Department of Orthopaedics, Faculty of Medicine, Juntendo University, 2-1-1, Hongo, Bunkyo-ku, Tokyo 113-0033, Japan
4. INSERM U955, IMRB Université Paris-Est, 51 avenue du Maréchal de Lattre de Tassigny, 94000 Créteil, France
5. Service de Chirurgie Orthopédique et Traumatologique, Hôpital Henri Mondor AP-HP, CHU Paris 12, Université Paris-Est, 51 avenue du Maréchal de Lattre de Tassigny, 94000 Créteil, France
6. Service de Chirurgie Maxillo-faciale, Centre Hospitalier Intercommunal Lucie & Raymond Aubrac , 40 Allée de la Source 94195 Villeneuve-Saint-Georges Cedex, France

* Corresponding author:

Guillaume Haïat

Laboratoire Modelisation et Simulation Multi Echelle, Centre National de la Recherche Scientifique, MSME UMR 8208 CNRS, 61 Avenue du General de Gaulle, 94010 Creteil, France.

Guillaume.haiat@cnrs.fr

**Abstract**

Summers osteotomy is a technique used to increase bone height and to improve bone density in dental implant surgery. The two main risks of this surgery, which is done by impacting an osteotome in bone tissue, are i) to perforate the sinus membrane and ii) the occurrence of benign paroxysmal vertigo, which are both related to excessive impacts during the osteotomy. Therefore, impacts must be carefully modulated. The aim of this study is to determine whether an instrumented hammer can predict bone damage before the total osteotome protrusion. 35 osteotomies were performed in 9 lamb palate samples using a hammer instrumented with a force sensor to record the variation of the force as a function of time $s(t)$. A signal processing was developed to determine the parameter $\tau$ corresponding to the time between the first two peaks of $s(t)$. A camera was used to determine the impact number for damage: $N_{Video}$. The surgeon determined when damage occurred, leading to $N_{Surg}$. An algorithm was developed to detect bone damage based on the variation of $\tau$ as a function of the impact number, leading to $N_{crit}$. The algorithm was always able to detect bone damage before total protrusion of the osteotome. We obtained $N_{Video} - N_{Crit} > -2$ (respectively $N_{Surg} - N_{Crit} > -2$) for 97 % (respectively 94 %) of the cases, which indicates the algorithm was almost always able to detect bone damage at most one impact after the video (respectively the surgeon). Our results pave the way to safer Summers osteotomy.

1. **Introduction**

Dental implants have been widely used to restore missing teeth. Dental implants placed in the posterior maxillary bone may be difficult when bone height and/or quality is not sufficient. An increase of the maxillary sinus volume coupled with bone resorption, related to tooth extraction or aging, may cause a reduction in the quality of the alveolar bone in which the implants are placed [1]. Therefore, before dental implant surgery, it is sometimes necessary to raise the sinus floor and add bone substitute biomaterial (BSB) to enhance bone regeneration and obtain a sufficient anchorage, in order to allow a proper dental implant primary stability.

Summers osteotomy is a technique used to increase bone height and improve bone quality when the patient has a subsinus bone height greater than 5 mm [1]. During this surgery, an osteotome is impacted using a mallet to drill the maxillary bone in order to introduce the BSB. One of the main danger lies in the perforation of the sinus membrane, also called Schneider membrane, located between the bone and the maxillary sinus. Since it is difficult to quantify the progress of the osteotome, surgeons still rely on their proprioception to decide when to stop the impacts on the osteotome [2]. The challenge lies in obtaining a sufficiently large cavity to allow the introduction of BSB, without perforating the sinus membrane. In the absence of visual assessment of the sinus membrane, it remains difficult to confirm that it was not damaged during the fracture of the maxillary bone [3]. If the cavity is too large or if the membrane is damaged, the filler may leak into the sinus cavity, leading to the impossibility of inserting the implant [4] and to a healing period of at least 3 months before revenging [5]. Moreover, the per operatory diagnosis of the perforation of the sinus membrane remains difficult.

Another challenge with Summers osteotomy lies in the occurrence of benign paroxysmal vertigo [6,7], which have been associated to excessive impacts of the osteotome while performing such osteotomies. Therefore, impacts must be realized very carefully and a compromise should be found considering the impact energy between a sufficiently high amplitude to allow the creation of the cavity and a sufficiently low amplitude to avoid benign paroxysmal vertigo [8].

As a consequence, a quantitative method allowing to detect the perforation of the osteotome would be very useful to help the surgeon perform this surgical procedure more safely. Several alternative techniques such as the trephine osteotomy and the novel bone graft delivery system have been introduced [9,10] but do not allow to detect when the osteotome induces a bone fracture. Other techniques have been developed to cut the bone, such as piezosurgery [11], which may help preserving the membrane, as it does not cut soft tissue. Although very useful, this technique requires additional equipment and increases procedure time.

In recent years, our group has developed an approach based on impact analyses allowing to assess the biomechanical properties of i) bone-implant systems and ii) the bone tissue located around an osteotome. The principle of the measurement is based on the use of a hammer instrumented with a piezoelectric force sensor, which measures the variation of the force as a function of time during each impact applied to the osteotome. The technique was initially applied to hip implant insertion *in vitro* [12–14], *ex vivo* [15,16] and then with anatomical subjects [17,18]. Static [19,20] and dynamic [21,22] finite element models were developed to better understand the mechanical phenomena that occur during hip implant insertion. Moreover, this instrumented hammer was tested in the context of osteotomy procedures. An *in vitro* study showed that it is possible to assess the plate thickness and material properties [23] as well as the progression of an osteotome in a bone sample [24]. An animal model was used to show that the instrumented hammer can detect when the osteotome approaches the end of the sample [24] and to assess the osteotome displacement [25]. Eventually, a study on anatomical subjects proved the interest of the technique in the context of osteotomies performed during rhinoplasty [26,27].

Based on the aforementioned feasibility studies, the aim of the present study is to determine whether an instrumented hammer can predict bone damage before the total protrusion of the osteotome in the context of Summers osteotomy. To do so, our approach is to consider a lamb palate animal model.

## 2. Materials and Methods
### a. Sample Preparation

A lamb model was chosen because it is commonly used to for osteotomies in maxillofacial surgery [28,29]. Nine lamb heads were collected from the local butcher's store and kept frozen prior to the experiments, similar to our previous study [16]. This protocol was approved by the local ethical committee of the Université Paris-Est Creteil.

The lamb palates were extracted from the remaining skull by following several steps. First, the mandible was separated from the skull using a power saw. Second, the upper jaw behind the teeth was removed from the back of the skull. Eventually, the palate was separated at the nostrils from the head. The raphe palatine, located in the center of the palate was cut (see Fig. 1) in order to allow a proper view of the osteotome. All soft tissues were removed on both sides of the palate for the same reason.

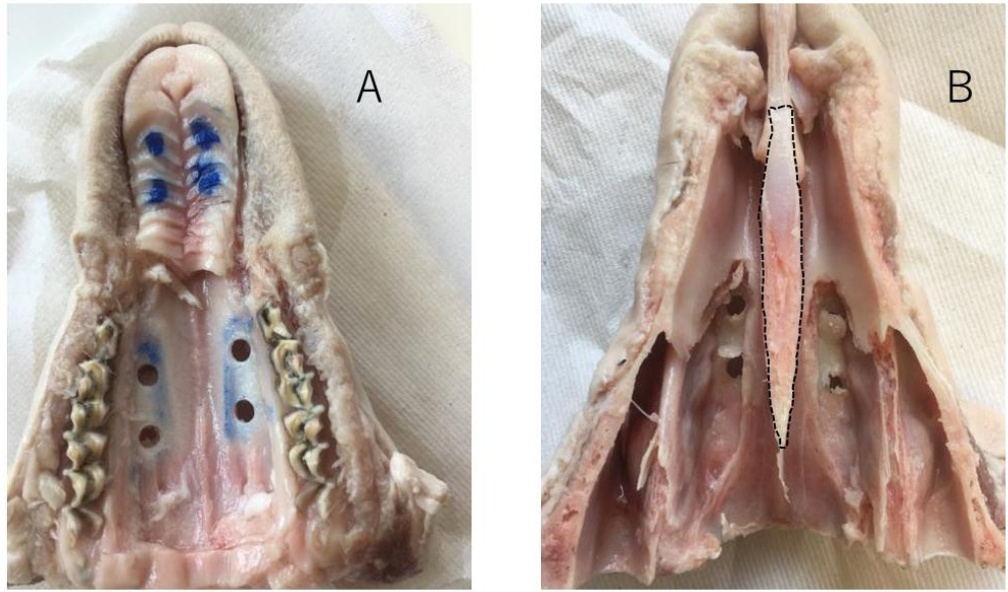

Figure. 1: Image of the sample corresponding to the lamb's palate after the osteotomies were performed. A: view from the caudal side. B: view from the sinus after the osteotomy. The raphe palatine (indicated by black dashed lines), located in the center of the palate, was cut.

b. Instrumented hammer and osteotome

In this study, a 260 g surgical mallet (32-6906-26, Zepf, Tuttlingen, Germany) was used to impact an osteotome (354.05, Acteon, Mérignac, France) with a diameter of 5 mm, which was held manually, similarly as what is done in the clinic. A dynamic piezoelectric force sensor (208C04, PCB Piezotronics, Depew, NY, USA) with a compression measurement range of up to 4.45 kN was screwed into the center of the impacting face of the hammer to measure the variation of the force applied to the osteotome as a function of time. A data acquisition module (NI 9234, National Instruments, Austin, TX, USA) with a sampling frequency of 51.2 kHz and a resolution of 24 bits was used to record the time variation of the force exerted on the osteotome. Data were transferred to a computer and recorded with a LabVIEW interface (National Instruments, Austin, TX, USA) for a duration of 20 ms. This duration was chosen as a compromise between i) a sufficiently high value to make sure to record all interesting information in all cases and ii) a sufficiently low value to minimize data acquisition duration for future real-time use. A schematic description of the experimental setup is shown in Fig. 2.

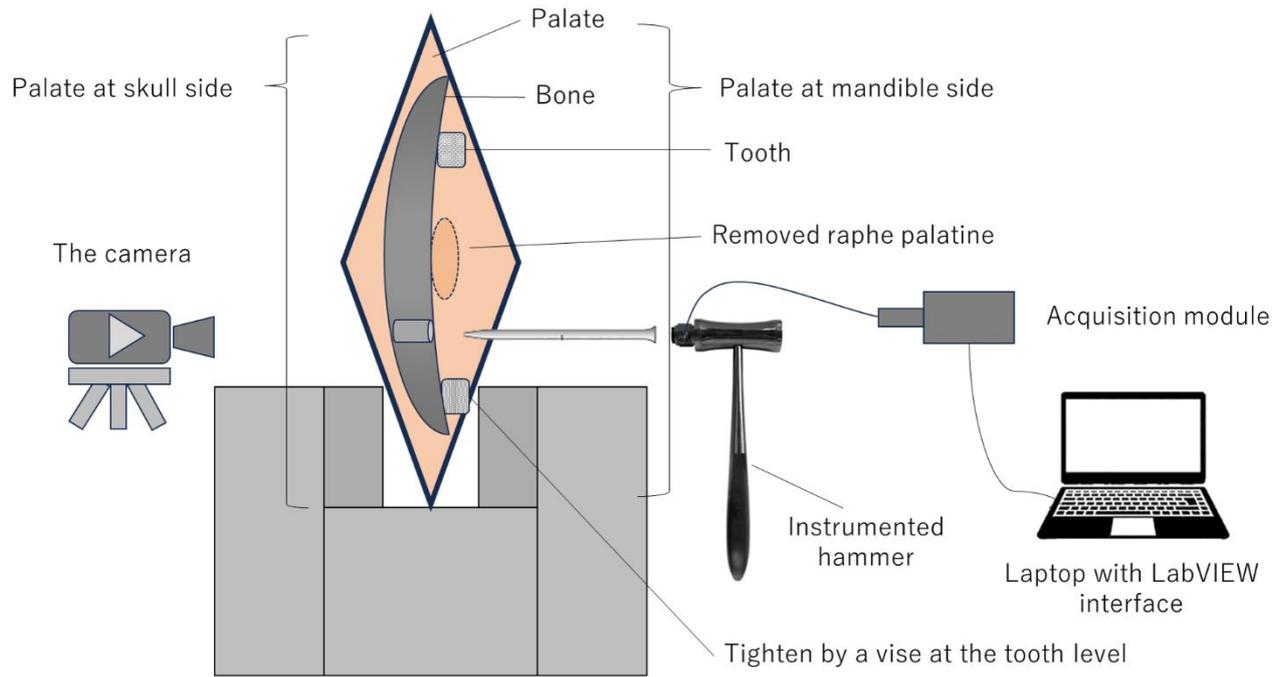

Figure. 2: Schematic description of the experimental setup.

    c.    Experimental Procedure

Four osteotomies were performed on each sample by an experienced maxillofacial surgeon (see Fig. 1), leading to a total number of 35 osteotomies and 811 impacts. When performing a Summers osteotomy in a clinical setting, a hole is drilled through the cortical bone until the thickness of the remaining bone tissue is equal to around 1 mm. Since it was not possible to measure the thickness of the remaining bone, we used an electric motor (Zimmer, Palm Beach Gardens, FL, USA) to drill holes until the palate was thin enough so that the surgeon finger could be detected by transparency. The palate was then placed horizontally in a vise and tightened at the tooth level, as described in Fig. 2. The surgeon was located on the side of the jaw and the camera was placed on the opposite side, in front of the sinus. The camera (L-920M3, Spedal, Taiwan) was located close to the sinus to detect when the osteotome emerges from the bone surface. Figure 3 shows images of each stage of the progression of the osteotome, which were extracted from the video.

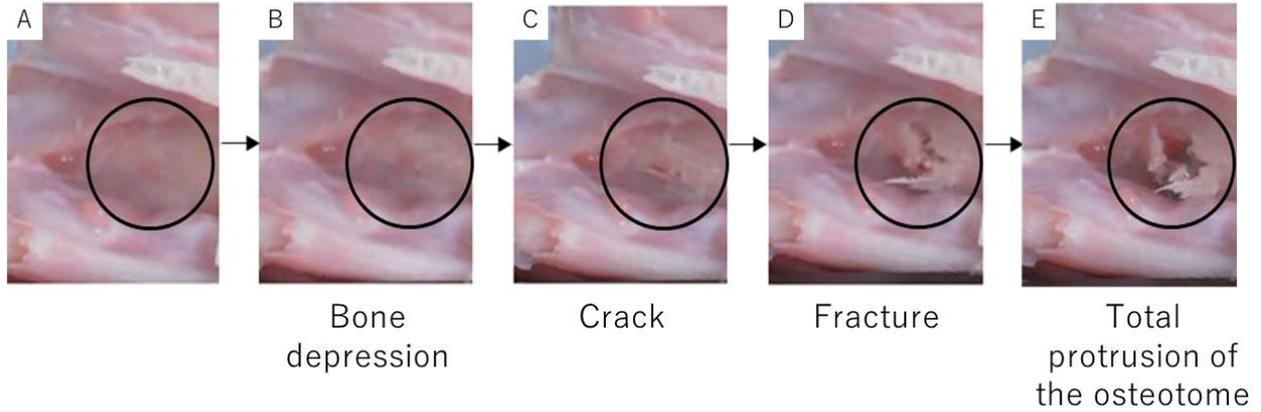

Figure. 3: Images of the sample obtained by the video analysis during the same osteotomy procedure corresponding to the displacement of the osteotome. A: the sample is intact. B: a bone depression occurs. C: A crack is initiated. D: A fracture occurs. E: The osteotome crosses the bone and is visible.

During the osteotomy procedure, the surgeon used the instrumented hammer to impact the osteotome, which was placed in the pre-drilled holes. The surgeon worked blindly, without access to the images of the camera nor to the results obtained with the instrumented hammer in order to avoid being influenced by any other feedback than his proprioception, which corresponds to the clinical situation. During the experimental procedure, the impacts were numbered successively. The "surgeon damage", noted $N_{Surg}$, corresponds the impact number for which the surgeon thinks damage occurs. The "video fracture", noted $N_{Video}$, refers to the impact number for which a fracture is visible on the video (see Fig. 3D). Eventually, "total protrusion", noted $N_{Prot}$, corresponds to the impact number when the osteotome has completely crossed the bone and is fully visible on the video (see Fig. 3E).

d. Data processing

Data processing was performed following the same method described in [23,24] and briefly recalled in what follows. The time dependence of the force applied to the osteotome during a given impact was measured with the force sensor described above, leading to a signal denoted *s(t)*. An example of a typical signal is shown in Fig. 4. Several peaks can be observed in this force signal. As shown in [23], the first peak (see first contact in Fig.4) corresponds to the initial impact of the hammer on the osteotome, while the following peaks correspond to the rebound of the osteotome between the hammer and the osteotomized material. Each peak corresponds to a contact between the instrumented hammer and the osteotome. Note that a 1-D analytical model [23] where the osteotome and the bone were modeled by spring was successfully used to simulate typical signal corresponding to the one shown in Fig. 4. A dedicated signal processing technique was applied to *s(t)* using information derived from the different peaks obtained in the signal. The time of maximum of the first two peaks of *s(t)* was determined for each impact. The indicator $\tau$ was then defined as the difference between the times of the second and first peaks of *s(t)*.

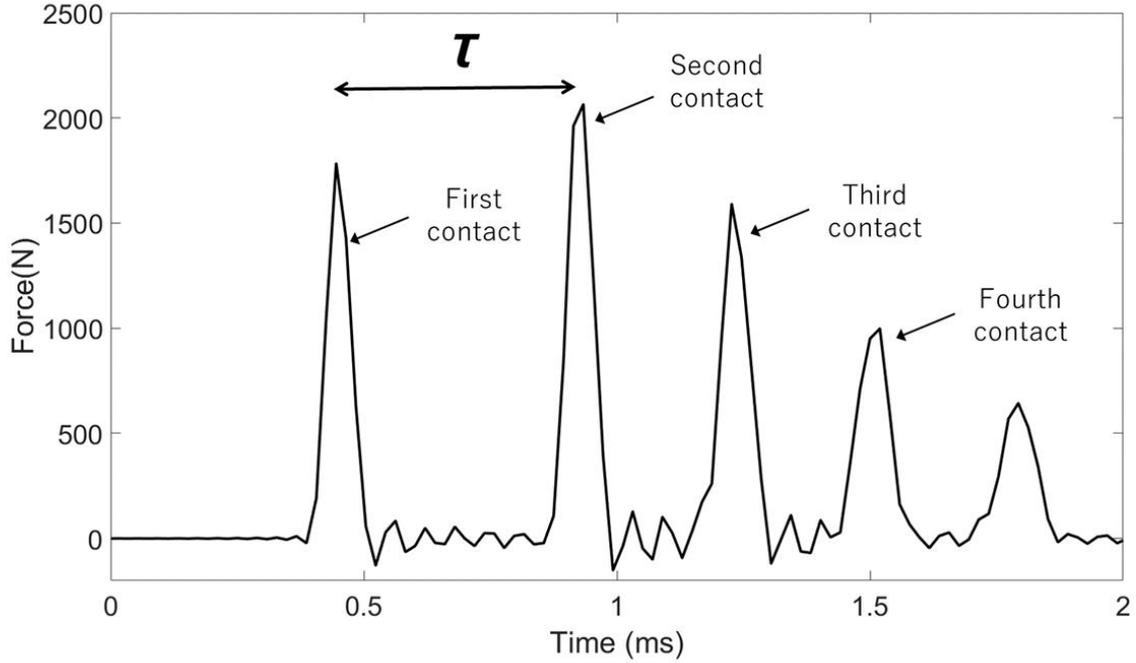

Figure. 4: Example of a typical signal *s(t)* corresponding to the variation of the force as a function of time obtained during an impact of the instrumented hammer on the osteotome.

e. Damage criterion

Following the method developed in [24], a criterion based on the variation of $\tau$ as a function of the impact number during each osteotomy procedure was developed in order to determine when bone damage occurs. As demonstrated in [23] the indicator $\tau$ is related to the rigidity of the material surrounding the osteotome tip and thus an increase of $\tau$ is expected when the bone is damaged because cracks induce a decrease of the overall bone rigidity. This criterion is based on the comparison between the value of $\tau(n)$ for the impact considered #*n* (corresponding to the impact analyzed by the algorithm) and the average value of $\tau$ obtained for previous impacts. An increase of the value of $\tau$ is considered to be associated with bone damage and hence to a decrease of the local bone stiffness. When possible, the last 15 impacts are considered (see Ineq. 1), while in the case where the path has fewer than 15 impacts, the criterion takes into account all impacts before (see Ineq. 2). Let $\alpha$ be a real number that will be determined during the optimization procedure (see below). The detection of bone damage is considered at the impact number n=$N(\alpha)$ when one of the first following conditions is met:

1. If the osteotomy has more than 15 impacts ($n > 15$):

$$\tau(n) > \alpha * \frac{1}{15} * \sum_{j=n-15}^{n-1} \tau(j) \qquad \text{(Ineq.1)}$$

2. If the osteotomy has less than 15 impacts ($n \leq 15$):

$$\tau(n) > \alpha * \frac{1}{n-1} * \sum_{j=1}^{n-1} \tau(j) \qquad \text{(Ineq. 2)}$$

The optimal value of $\alpha$ (denoted $\alpha_{optimal}$) is determined following the optimization procedure described below and this optimal α value will lead to a parameter called "$N_{Crit}$" referring to the prediction of the instrumented mallet for $N$ ($\alpha = \alpha_{optimal}$). The goal was to determine the optimal value of $\alpha$ allowing a minimization of $D$, which is defined by the average value of $|N(\alpha) - N_{Video}|$ over the entire database with the additional constraint of avoiding a situation where $N(\alpha) > N_{Prot}$ (to make sure that the surgeon is always warned before total protrusion). We note $M$ the number of samples for which $N(\alpha) > N_{Prot}$. Hence, the optimization algorithm consists of minimizing the value of $D$ with the constraint of having $M=0$.

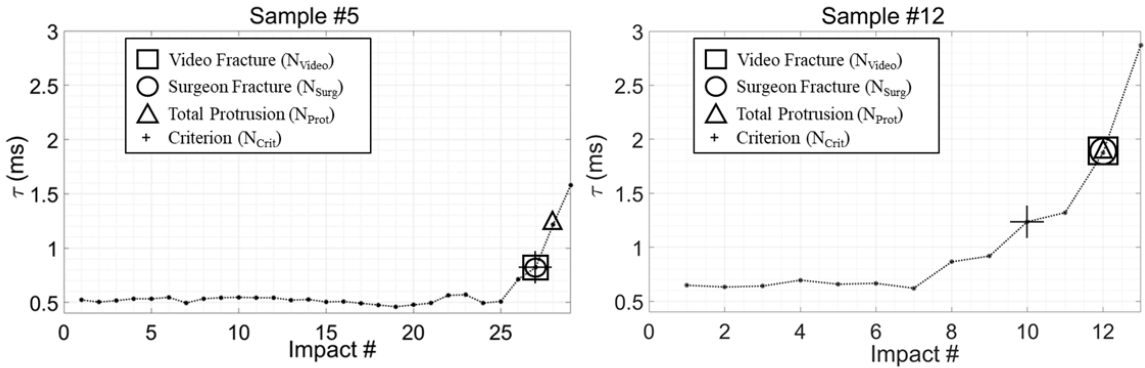

Figure. 5: Example of the variation of $\tau$ as a function of the impact number for a given osteotomy. The impact numbers corresponding to the prediction of the instrumented hammer ($N_{Crit}$) total protrusion ($N_{Prot}$), surgeon damage ($N_{Surg}$) and video fracture ($N_{Video}$) are indicated with different symbols. A (sample #5): the instrumented hammer detects damage at the same time as the video and the surgeon. B (sample #12): the instrumented hammer detects damage two impacts before the video and the surgeon, thus warning the surgeon of the upcoming fracture.

## 3. Results

Figure 5 shows the variation of $\tau$ as a function of the impact number for the sample #5 (left) and #12 (right). The video fracture, the surgeon damage and the total protrusion of the osteotome are also indicated. First, the values of $\tau$ formed a plateau at about 0.518 ± 0.028 ms (left) and 0.652 ± 0.025 (right), which increases slightly after impact #26 (left) and #8 (right). Finally, $\tau$ increased significantly to reach 1.218 ms (left) and 1.874 ms (right) at the total protrusion of the osteotome.

Statistical parameters corresponding to the values of $\tau$ obtained for the 35 osteotomies procedures are shown in Table 1. The mean values of $\tau$ (ms) after the fracture predicted with the video and by the surgeon are about 1 ms higher than beforehand.

| $\tau$ (ms) | Video Fracture ($N_{Video}$) | | Surgeon Damage ($N_{Surg}$) | | Total Protrusion ($N_{Prot}$) | | Damage criterion ($N_{Crit}$) | |
|---|---|---|---|---|---|---|---|---|
| Timing | Before | After | Before | After | Before | After | Before | After |
| Mean value | 0.637 | 1.671 | 0.633 | 1.591 | 0.655 | 1.991 | 0.601 | 1.483 |
| Standard deviation | 0.192 | 1.036 | 0.189 | 1.015 | 0.220 | 1.115 | 0.136 | 0.896 |

Table. 1: Average and standard deviation values of $\tau$ obtained before and after the impact corresponding to damage predicted by the instrumented hammer ($N_{Crit}$), by the video ($N_{Video}$) and by the surgeon ($N_{Surg}$).

Figure 6 shows the variation of $D$ and $M$ as a function of $\alpha$ (see subsection 2.5). The value of $D$ is shown to decrease as a function of $\alpha$ until $\alpha = 1.392$, while $M = 0$ for $\alpha < 1.392$. The results shown in Fig. 6 explain why the optimization algorithm leads to a value of $\alpha = 1.392$. In what follows, let us note $N_{crit} = N(1.392)$, which corresponds to the optimal configuration. $N_{Crit}$ is indicated in Fig. 5 by crosses and corresponds to the final output of the prediction of the instrumented hammer.

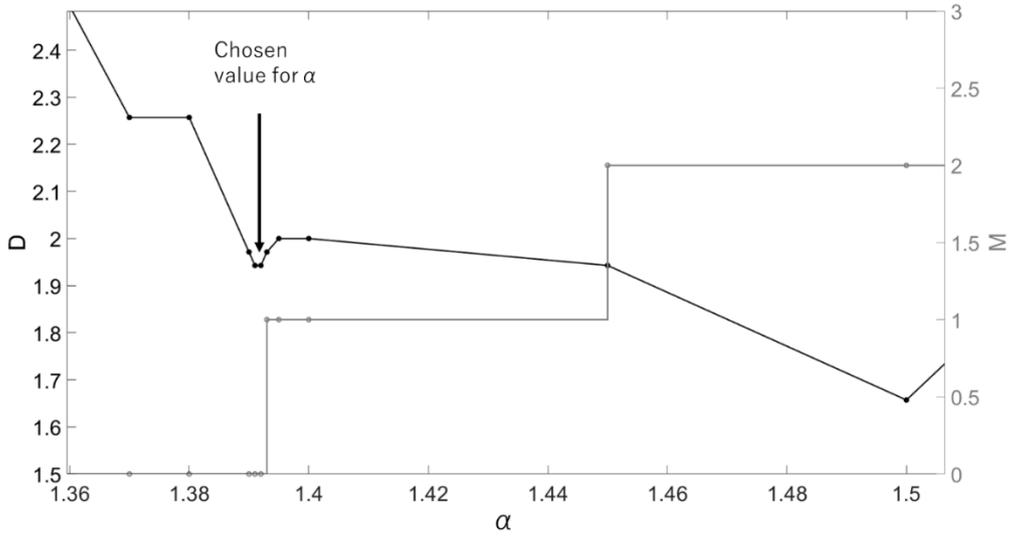

Figure. 6: Performance of the algorithm used by the instrumented hammer for different values of the coefficient $\alpha$. The black curve corresponds to the variation of the average difference between the video and the criterion detection (*i.e.* the average value of $|N(\alpha) - N_{Video}|$ over the entire database), called $D$, as a function of $\alpha$. The grey curve represents the number of samples for which the criterion detection occured after the total protrusion (*i.e.* $N(\alpha) > N_{Prot}$), called $M$, as a function of $\alpha$.

Figure 5 shows that for the sample #5, $N_{Crit} = N_{Surg} = N_{Video} = 27$, while for the sample #12, $N_{Crit} = 10$ and $N_{Surg} = N_{Video} = 12$, which indicates that the instrumented hammer is able to detect damage two impacts before the video and the surgeon.

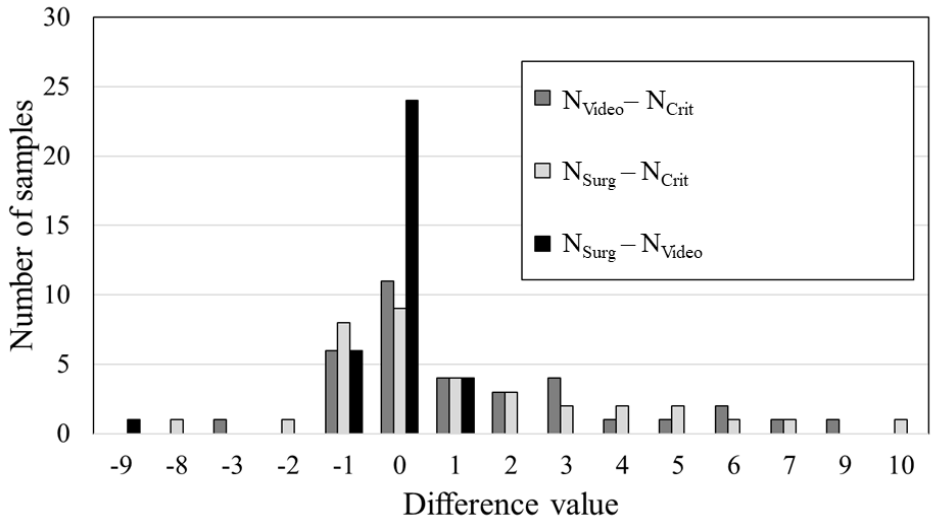

Figure. 7: Differences between the impact number corresponding to the damage predicted by the instrumented hammer ($N_{Crit}$), by the video ($N_{Video}$) and by the surgeon ($N_{Surg}$).

Figure 7 shows the histogram corresponding to the differences obtained between $N_{Crit}$, $N_{Surg}$ and $N_{Video}$ for all osteotomies pooled together. Note that since $M = 0$, $N_{Crit}$ was always lower than $N_{Prot}$. For all osteotomies except one, $|N_{Surg} - N_{Video}| \leq 1$, which indicates that the surgeon almost always agrees with the video, which will be discussed in section 4. For a given osteotomy, we obtained $N_{Surg} - N_{Video} = -9$, which indicates that the surgeon was quiet early in detecting damage. Note for this same osteotomy, we had $N_{Video} - N_{Crit} = 1$, which indicates that the instrumented hammer was in better agreement with the video than the surgeon. For all other osteotomies except one (respectively two), $N_{Video} - N_{Crit} > -2$ (respectively $N_{Surg} - N_{Crit} > -2$), which indicates that the instrumented hammer was almost always able to detect damage either one impact after, at the same time or before the video and the surgeon. Note that in several cases, the instrumented hammer was able to detect damage before the video and the surgeon ($N_{Video} - N_{Crit} > 0$ and $N_{Surg} - N_{Crit} > 0$).

## 4. Discussion

Although Summers osteotomy is widely performed in the clinic to increase bone height and improve bone density for the implantation of dental implants, this surgical technique is largely dependent on the surgeon proprioception. Therefore, providing the surgeon a decision-support system to allow a safer surgical procedure is expected to lead to a major technological breakthrough and improvement of such surgical procedure. The originality of the present *ex vivo* study is to demonstrate that an instrumented hammer can be used to predict damage of the palate bone before total protrusion of the osteotome in the context of Summers osteotomy. The results obtained herein are in agreement with our previous study using various materials, which demonstrated that the indictor $\tau$ i) was sensitive to the distance from the end of the bone sample [24] and ii) could predict the rupture of the plate samples [24].

### a. Physical interpretation

The physical interpretation of the behaviour of $\tau$ had already been addressed in previous experimental [12–14,19,20] and numerical studies [22,23]. The force applied by the hammer to the osteotome during the first contact induces an acceleration of the osteotome, which then bounces back and forth between bone tissue and the head of the hammer. This phenomenon leads to the different peaks of the force shown in Figure 4. Therefore, the time between the first and second peak (denoted $\tau$ herein) is related to the rigidity of the material around the osteotome tip. In Summers osteotomy, as shown in Table 1, $\tau$ increases significantly after damage is detected, because while the tip of the osteotome is still located in bone tissue before damage, the overall sample rigidity decreases significantly after damage due to the presence of microcracks, resulting in an increase of $\tau$. This behavior is in agreement with the results using plywood bone sample from [24], where a plateau was also followed by a sudden increase of $\tau$ before, during and after the rupture of the bone sample with the osteotome.

### b. Choices regarding the detection method

First, the connection between the hammer and the sensor is made by a stud, and is therefore assumed to be perfect. Transient imperfect contact between the sensor and the osteotome does occur on a regular basis since the contact is unilateral in this area, as described in subsection 2.d.

Second, the reproducibility of the estimation of $\tau$ by the instrumented hammer has been assessed at approximately 0.0020 ms by considering a dedicated configuration without modification of the position of the osteotome (data not shown). Moreover, an experienced surgeon performed the experiment in this study. The skill of hammering may be different among surgeons. In order to check that the user has no effect on the signal, two users were asked to perform five impacts on the same experimental configuration. The signals were similar and no significant difference between the values of $\tau$ measured for both users was obtained (data not shown). These results indicate that only $N_{Surg}$ will change when considering other users.

Third, in this study, we chose to only consider the indicator $\tau$ to detect bone fracture. In some previous studies [23,30], a parameter called $\lambda$, corresponding to the ratio between the amplitude of the first two peaks has been studied. Other information may be enclosed in the force signal and in particular in the other peaks shown in Fig. 4 such as for instance the time difference between each peak. Such information could provide more insight on the tissue properties. However, this point was not investigated herein and is left for future studies. In order to create a detection method that is robust and not user-dependent, the impact force value was not included in the analyses, as it is an operator-dependent parameter.

Fourth, we did not consider the use of a modal hammer, which gave interesting results in the context of hip arthroplasty [31], because it complicates the experimental set-up in the context of a future clinical use. It could lead to an increase of the operating time and create issues regarding the sterility of the instruments (osteotome) directly in contact with the patient.

### c. Damage criterion

The damage criterion showed relatively high accuracy compared to the two other techniques. Although a perfect match between the damage detection and the video fracture was observed in 31.4 %, the difference was within ±1 impacts in 60 % of the cases. Moreover, in 80% of the cases, damage detection could be done earlier or at the same time with the instrumented hammer than with the video, which can be explained by potential damages that cannot be seen with the video, *e.g.* microcracks may be present in the sample before they may be detected visually or using the surgeon proprioception. More importantly, the instrumented hammer could always detect damage before total protrusion, which is important to allow a safer surgery because the most dangerous situation is associated with excessive penetration of the osteotome in the sinus, which may lead to a rupture of its membrane.

The number of 15 impacts considered for the criterion (see Ineq. 1&2) was chosen as it corresponds to the bottom limit of the average number of impacts per osteotomy, *i.e.* to the mean minus the standard deviation of the number of impacts per sample. Please note that the number of impacts chosen for the average in the criterion has an effect on the values of *α* chosen after the optimization.

Finally, further improvements can be brought to our approach. First, it would be interesting to consider additional samples to increase the size of the database and refine the result. Second, the implementation of a numerical model could allow to better understand and control the experimental parameters, and thus look for other indicators in the signal. Eventually, the implementation of an artificial intelligence algorithm could allow to create a more efficient criterion, similarly as what was done in [30].

d. Limitations

This study has several limitations.

Three complementary methods were used to detect damage, each method having potential associated errors. First, while the surgeon is experienced, his proprioception cannot be perfectly accurate and potential errors are always possible because it is an empirical approach. Second, video analysis does not provide a volume information and therefore does not allow a quantitative assessment of potential damage in the bone volume, particularly when the cracks are not open. Third, the instrumented hammer also has limitation related to the fact that this technique is new and currently under development. Despite these limitations, an overall good agreement is obtained between the different techniques. Note that the agreement between the surgeon and the video is excellent, which may be explained by the fact that i) the surgeon has visual access to the bone, ii) all soft tissues were removed and iii) the sample was rigidly fixed in a vise. Therefore, the present results must be confirmed in the future in anatomical subjects, which is a situation closer to the clinics. However, it was mandatory to carry out the experiments *ex vivo* before considering anatomical subjects because of ethical constraints.

Second, there are significant differences between the clinical situation and the present animal model, in particular in terms of general configuration (fixation of the sample in a vice), of bone properties and of the presence of soft tissues. Note that the presence of soft tissues was shown not to affect the results in a previous paper [32] considering the use of the instrumented hammer in total hip replacement, which needs however to be confirmed in the context of Summers osteotomy. Note that the approach considering an animal model is similar to previous works done for the development of the instrumented hammer in the context of rhinoplasty [25,26].

Third, only one osteotome was used, while there are many commercially available osteotomes on the market, which may affect the results [33]. The size and shape of the osteotome may modify the signals obtained with the instrumented hammer, which explains why other osteotome geometry should also be tested.

## 5. Conclusion

This study is the first to evaluate the performance of an instrumented hammer to assess damage while performing Summers osteotomy in an animal model. The results show the performance of the method, which is often able to predict bone damage before the video and the surgeon proprioception. Moreover, our approach always allows to warn the surgeon before the osteotome fully crosses bone tissue, which consists in a valuable information. However, this technique must be validated in anatomical subjects before performing future clinical trials. Our results pave the path to the development of a decision-support system for the surgeon, which will allow a safer surgery using objective parameters in real time.

## 6. Acknowledgements

This project has received funding from the project OrthAncil (ANR-21- CE19-0035-03), from the project OrthoMat (ANR-21-CE17-0004), from the project ModyBe (ANR-23-CE45-0011-02) and from the project MaxillOsteo (ANR-24-CE17-2236)). This project has received funding from the European Research Council (ERC) under Horizon 2020 (grant agreement # 101062467, project ERC Proof of Concept Impactor).

Competing interests: None declared
This protocol was approved by the local ethical committee of the Université Paris-Est Creteil.

**Reference**

1. Antoun, H.; El-Zoghbi, H.; Cherfane, P.; Missika, P. Les Ostéotomes de Summers : Une Alternative Au Volet Latéral Pour Les Soulevés de Sinus ? *Implantodontie* **2003**, *12*, 3–9, doi:10.1016/j.implan.2003.10.003.

2. Fermergård, R.; Åstrand, P. Osteotome Sinus Floor Elevation without Bone Grafts – A 3-Year Retrospective Study with Astra Tech Implants. *Clin Implant Dent Rel Res* **2012**, *14*, 198–205, doi:10.1111/j.1708-8208.2009.00254.x.

3. Díaz-Olivares, L.A.; Cortés-Bretón Brinkmann, J.; Martínez-Rodríguez, N.; Martínez-González, J.M.; López-Quiles, J.; Leco-Berrocal, I.; Meniz-García, C. Management of Schneiderian Membrane Perforations during Maxillary Sinus Floor Augmentation with Lateral Approach in Relation to Subsequent Implant Survival Rates: A Systematic Review and Meta-Analysis. *Int J Implant Dent* **2021**, *7*, 91, doi:10.1186/s40729-021-00346-7.

4. Viña-Almunia, J.; Peñarrocha-Diago, M.; Peñarrocha-Diago, M. Influence of Perforation of the Sinus Membrane on the Survival Rate of Implants Placed after Direct Sinus Lift. Literature Update. *Med Oral Patol Oral Cir Bucal* **2009**, *14*, E133-136.

5. Ferrigno, N.; Laureti, M.; Fanali, S. Dental Implants Placement in Conjunction with Osteotome Sinus Floor Elevation: A 12-Year Life-Table Analysis from a Prospective Study on 588 ITI Implants. *Clinical oral implants research* **2006**, *17*, 194–205, doi:10.1111/j.1600-0501.2005.01192.x.

6. Giannini, S.; Signorini, L.; Bonanome, L.; Severino, M.; Corpaci, F.; Cielo, A. Benign Paroxysmal Positional Vertigo (BPPV): It May Occur after Dental Implantology. A Mini Topical Review. *Eur Rev Med Pharmacol Sci* **2015**, *19*, 3543–3547.

7. Peñarrocha-Diago, M.; Rambla-Ferrer, J.; Perez, V.; Pérez-Garrigues, H. Benign Paroxysmal Vertigo Secondary to Placement of Maxillary Implants Using the Alveolar Expansion Technique with Osteotomes: A Study of 4 Cases. *Int J Oral Maxillofac Implants* **2008**, *23*, 129–132.

8. Molina, A.; Sanz-Sánchez, I.; Sanz-Martín, I.; Ortiz-Vigón, A.; Sanz, M. Complications in Sinus Lifting Procedures: Classification and Management. *Periodontol 2000* **2022**, *88*, 103–115, doi:10.1111/prd.12414.

9. Emtiaz, S.; Caramês, J.M.M.; Pragosa, A. An Alternative Sinus Floor Elevation Procedure: Trephine Osteotomy. *Implant Dentistry* **2006**, *15*, 171–177, doi:10.1097/01.id.0000220550.27164.74.

10. Mazor, Z.; Ioannou, A.; Venkataraman, N.; Kotsakis, G. A Minimally Invasive Sinus Augmentation Technique Using a Novel Bone Graft Delivery System. *International Journal of Oral Implantology & Clinical Research* **2013**, *4*, 78–82, doi:10.5005/jp-journals-10012-1097.

11. Mazor, Z.; Kfir, E.; Lorean, A.; Mijiritsky, E.; Horowitz, R.A. Flapless Approach to Maxillary Sinus Augmentation Using Minimally Invasive Antral Membrane Balloon Elevation. *Implant Dentistry* **2011**, *20*, 434–438, doi:10.1097/ID.0b013e3182391fe3.

12. Mathieu, V.; Michel, A.; Lachaniette, C.-H.F.; Poignard, A.; Hernigou, P.; Allain, J.; Haïat, G. Variation of the Impact Duration during the in Vitro Insertion of Acetabular Cup Implants. *Medical Engineering & Physics* **2013**, *35*, 1558–1563, doi:10.1016/j.medengphy.2013.04.005.

13. Michel, A.; Bosc, R.; Mathieu, V.; Hernigou, P.; Haiat, G. Monitoring the Press-Fit Insertion of an Acetabular

Cup by Impact Measurements: Influence of Bone Abrasion. *Proc Inst Mech Eng H* **2014**, *228*, 1027–1034, doi:10.1177/0954411914552433.

14. Michel, A.; Bosc, R.; Vayron, R.; Haiat, G. In Vitro Evaluation of the Acetabular Cup Primary Stability by Impact Analysis. *Journal of Biomechanical Engineering* **2015**, *137*, doi:10.1115/1.4029505.

15. Lomami, H.A.; Damour, C.; Rosi, G.; Poudrel, A.-S.; Dubory, A.; Flouzat-Lachaniette, C.-H.; Haiat, G. Ex Vivo Estimation of Cementless Femoral Stem Stability Using an Instrumented Hammer. *Clinical Biomechanics* **2020**, *76*, doi:10.1016/j.clinbiomech.2020.105006.

16. Michel, A.; Bosc, R.; Sailhan, F.; Vayron, R.; Haiat, G. Ex Vivo Estimation of Cementless Acetabular Cup Stability Using an Impact Hammer. *Medical Engineering & Physics* **2016**, *38*, 80–86, doi:10.1016/j.medengphy.2015.10.006.

17. Michel, A.; Bosc, R.; Meningaud, J.-P.; Hernigou, P.; Haiat, G. Assessing the Acetabular Cup Implant Primary Stability by Impact Analyses: A Cadaveric Study. *PLOS ONE* **2016**, *11*, doi:10.1371/journal.pone.0166778.

18. Dubory, A.; Rosi, G.; Tijou, A.; Lomami, H.A.; Flouzat-Lachaniette, C.-H.; Haïat, G. A Cadaveric Validation of a Method Based on Impact Analysis to Monitor the Femoral Stem Insertion. *Journal of the Mechanical Behavior of Biomedical Materials* **2020**, *103*, doi:10.1016/j.jmbbm.2019.103535.

19. Nguyen, V.-H.; Rosi, G.; Naili, S.; Michel, A.; Raffa, M.-L.; Bosc, R.; Meningaud, J.-P.; Chappard, C.; Takano, N.; Haiat, G. Influence of Anisotropic Bone Properties on the Biomechanical Behavior of the Acetabular Cup Implant: A Multiscale Finite Element Study. *Computer Methods in Biomechanics and Biomedical Engineering* **2017**, *20*, 1312–1325, doi:10.1080/10255842.2017.1357703.

20. Raffa, M.L.; Nguyen, V.-H.; Tabor, E.; Immel, K.; Housset, V.; Flouzat-Lachaniette, C.-H.; Haiat, G. Dependence of the Primary Stability of Cementless Acetabular Cup Implants on the Biomechanical Environment. *Proc Inst Mech Eng H* **2019**, *233*, 1237–1249, doi:10.1177/0954411919879250.

21. Michel, A.; Nguyen, V.-H.; Bosc, R.; Vayron, R.; Hernigou, P.; Naili, S.; Haiat, G. Finite Element Model of the Impaction of a Press-Fitted Acetabular Cup. *Medical & Biological Engineering & Computing* **2016**, *55*, 781–791, doi:10.1007/s11517-016-1545-2.

22. Poudrel, A.-S.; Bouffandeau, A.; Rosi, G.; Dubory, A.; Lachaniette, C.-H.F.; Nguyen, V.-H.; Haiat, G. 3-D Finite Element Model of the Impaction of a Press-Fitted Femoral Stem under Various Biomechanical Environments. *Computers in Biology and Medicine* **2024**, *174*, 108405, doi:10.1016/j.compbiomed.2024.108405.

23. Hubert, A.; Rosi, G.; Bosc, R.; Haiat, G. Using an Impact Hammer to Estimate Elastic Modulus and Thickness of a Sample During an Osteotomy. *Journal of Biomechanical Engineering* **2020**, *142*, doi:10.1115/1.4046200.

24. Bas dit Nugues, M.; Rosi, G.; Hériveaux, Y.; Haïat, G. Using an Instrumented Hammer to Predict the Rupture of Bone Samples Subject to an Osteotomy. *Sensors* **2023**, *23*, 2304, doi:10.3390/s23042304.

25. Lamassoure, L.; Giunta, J.; Rosi, G.; Poudrel, A.-S.; Bosc, R.; Haïat, G. Using an Impact Hammer to Perform Biomechanical Measurements during Osteotomies: Study of an Animal Model. *Proceedings of the Institution of Mechanical Engineers, Part H: Journal of Engineering in Medicine* **2021**, *235*, 838–845,

doi:10.1177/09544119211011824.

26. Lamassoure, L.; Giunta, J.; Rosi, G.; Poudrel, A.-S.; Meningaud, J.-P.; Bosc, R.; Haïat, G. Anatomical Subject Validation of an Instrumented Hammer Using Machine Learning for the Classification of Osteotomy Fracture in Rhinoplasty. *Medical engineering & Physics* **2021**, *95*, 111–116, doi:10.1016/j.medengphy.2021.08.004.

27. Giunta, J.; Lamassoure, L.; Nokovitch, L.; Rosi, G.; Poudrel, A.-S.; Meningaud, J.-P.; Haïat, G.; Bosc, R. Validation of an Instrumented Hammer for Rhinoplasty Osteotomies: A Cadaveric Study. *Facial Plastic Surgery & Aesthetic Medicine* **2021**, doi:10.1089/fpsam.2021.0107.

28. Valbonetti, L.; Berardinelli, P.; Scarano, A.; Piattelli, A.; Mattioli, M.; Barboni, B.; Vulpiani, M.P.; Muttini, A. Translational Value of Sheep as Animal Model to Study Sinus Augmentation. *Journal of Craniofacial Surgery* **2015**, *26*, 737–740, doi:10.1097/SCS.0000000000001785.

29. López-Niño, J.; García-Caballero, L.; González-Mosquera, A.; Seoane-Romero, J.; Varela-Centelles, P.; Seoane, J. Lamb Ex Vivo Model for Training in Maxillary Sinus Floor Elevation Surgery: A Comparative Study With Human Standards. *Journal of Periodontology* **2012**, *83*, 354–361, doi:10.1902/jop.2011.110210.

30. Bas Dit Nugues, M.; Lamassoure, L.; Rosi, G.; Flouzat-Lachaniette, C.H.; Khonsari, R.H.; Haiat, G. An Instrumented Hammer to Detect the Rupture of the Pterygoid Plates. *Ann Biomed Eng* **2024**, doi:10.1007/s10439-024-03596-9.

31. Poudrel, A.-S.; Rosi, G.; Nguyen, V.-H.; Haiat, G. Modal Analysis of the Ancillary During Femoral Stem Insertion: A Study on Bone Mimicking Phantoms. *Ann Biomed Eng* **2022**, *50*, 16–28, doi:10.1007/s10439-021-02887-9.

32. Bosc, R.; Tijou, A.; Rosi, G.; Nguyen, V.-H.; Meningaud, J.-P.; Hernigou, P.; Flouzat-Lachaniette, C.-H.; Haiat, G. Influence of Soft Tissue in the Assessment of the Primary Fixation of Acetabular Cup Implants Using Impact Analyses. *Clinical Biomechanics* **2018**, *55*, 7–13, doi:10.1016/j.clinbiomech.2018.03.013.

33. Kuran, I.; Özcan, H.; Usta, A.; Bas, L. Comparison of Four Different Types of Osteotomes for Lateral Osteotomy: A Cadaver Study. *Aesth. Plast. Surg.* **1996**, *20*, 323–326, doi:10.1007/bf00228464.

Figure legends

Figure. 1: Image of the sample corresponding to the lamb's palate after the osteotomies were performed. A: view from the caudal side. B: view from the sinus after the osteotomy. The raphe palatine (indicated by black dashed lines), located in the center of the palate, was cut.

Figure. 2: Schematic description of the experimental setup.

Figure. 3: Images of the sample obtained by the video analysis during the same osteotomy procedure corresponding to the displacement of the osteotome. A: the sample is intact. B: a bone depression occurs. C: A crack is initiated. D: A fracture occurs. E: The osteotome crosses the bone and is visible.

Figure. 4: Example of a typical signal $s(t)$ corresponding to the variation of the force as a function of time obtained during an impact of the instrumented hammer on the osteotome.

Figure. 5: Example of the variation of $\tau$ as a function of the impact number for a given osteotomy. The impact numbers corresponding to the prediction of the instrumented hammer ($N_{Crit}$) total protrusion ($N_{Prot}$), surgeon damage ($N_{Surg}$) and video fracture ($N_{Video}$) are indicated with different symbols. A (sample #5): the instrumented hammer detects damage at the same time as the video and the surgeon. B (sample #12): the instrumented hammer detects damage two impacts before the video and the surgeon, thus warning the surgeon of the coming fracture.

Figure. 6: Performance of the algorithm used by the instrumented hammer for different values of the coefficient $\alpha$. The black curve corresponds to the variation of the average difference between the video and the criterion detection (*i.e.* the average value of $|N(\alpha) - N_{Video}|$ over the entire database), called *D*, as a function of $\alpha$. The grey curve represents the number of samples for which the criterion detection occured after the total protrusion (*i.e.* $N(\alpha) > N_{Prot}$), called *M*, as a function of $\alpha$.

Figure. 7: Differences between the impact number corresponding to the damage predicted by the instrumented hammer ($N_{Crit}$), by the video ($N_{Video}$) and by the surgeon ($N_{Surg}$).

Table. 1: Average and standard deviation values of $\tau$ obtained before and after the impact corresponding to damage predicted by the instrumented hammer ($N_{Crit}$), by the video ($N_{Video}$) and by the surgeon ($N_{Surg}$).